\documentclass[preprint,showpacs,preprintnumbers,amsmath,amssymb]{revtex4}
\input psfig.sty
\usepackage{graphicx}% Include figure files
\usepackage{dcolumn}% Align table columns on decimal point
\usepackage{bm}% bold math

\begin{document}

\title{Negative heat capacity and non-extensive kinetic theory}

\author{R. Silva} \email{rsilva@uern.br}

\affiliation{Departamento de F\'{\i}sica, Universidade do Estado do Rio Grande do Norte,
59610-210, Mossor\'o - RN, Brazil}

\author{J. S. Alcaniz} \email{alcaniz@dfte.ufrn.br}

\affiliation{Astronomy Department, University of Washington, Seattle,
Washington, 98195-1580, USA}

\affiliation{Departamento de Fisica, Universidade Federal do Rio Grande do Norte, C.P. 1641, Natal - RN,
59072-970, Brazil
}

\date{\today}% It is always \today, today,
             %  but any date may be explicitly specified

\begin{abstract}
The negative nature of the heat capacity $C_V$ of
thermodynamically isolated self-gravitating systems is rediscussed
in the framework of a non-extensive kinetic theory. It is found
that the dependence of $C_V$  on the non-extensive parameter $q$
gives rise to a negative branch with the critical
value corresponding to $q = 5/3$ ($C_V\rightarrow - \infty$).

\end{abstract}

%\pacs{98.80.Es; 95.35.+d; 98.62.Sb}% PACS, the Physics and Astronomy
                             % Classification Scheme.
%\keywords{Suggested keywords}%Use showkeys class option if keyword
                              %display desired
\maketitle

Consider a simple analogy between
ideal gas and self-gravitating system. The temperature of this
system is defined by
\begin{equation}
{1\over 2}m v^2={3\over 2}k_B T,
\end{equation}
where $m$ is the mass of a particle (e.g., a star) and $k_B$ is
the Boltzmann's constant. If such a system is composed by $N$
particles, its total kinetic energy is $K={3\over2}Nk_BT$ which,
according to the virial theorem, is equal to minus its total
energy, i.e., $E=-K$. Therefore, the heat capacity of the system
is
\begin{equation}\label{1}
C_V={dE\over dT}=-{3\over 2}Nk_B,
\end{equation}
which is clearly a negative quantity. In other words, this means that by
losing energy the system automatically grows hotter, or still,
that a self-gravitating system  can never be in a true
thermodynamical equilibrium state (see \cite {lind} and
references therein). In
actual fact, it is well known that when discussed in the context of the
microcanonical ensemble a self-gravitanting system interacting
via long range forces exhibits a negative heat capacity
\cite{pad}. The negativeness of the heat capacity has been largely studied in
many astrophysical phenomena. For example, Lynden-Bell \& Wood
\cite{lind1} used such a concept to explain the gravothermal catastrophe
for globular clusters while Beckenstein \cite{beck} and Hawking \cite{haw}
used it in the context of black hole thermodynamics.

On the other hand, a considerable theoretical effort has been
concentrated on the study of the statistical description of a
large variety of physical systems, resulting in the extension of
the Boltzmann-Gibbs' statistical mechanics. Particular examples
are systems endowed with long duration memory, anomalous
diffusion, turbulence in pure-electron plasma, self-gravitating
systems or, more generally, systems endowed with long range
interactions. In order to deal with such problems, Tsallis \cite{T88}
proposed the following generalization of the Boltzmann-Gibbs (BG)
entropy formula
\begin{equation}
S_q = -k_B{1-\sum_{i} p_i^{q}\over (q-1)},
\end{equation}
where $p_i$ is the probability of the $i^{th}$ microstate and $q$
is a parameter quantifying the degree of non-extensivity. In the
limit $q\rightarrow 1$ the celebrated BG extensive formula,
$S_1=-k_B\sum_{i} p_i\ln p_i,$ is readily recovered as a
particular case.

In the astrophysical context, the first applications of this
non-extensive $q$-statistics studied stellar polytropes
\cite{plast} and the peculiar velocity function of galaxy clusters
\cite{lav}. The Jeans gravitational instability criterion for a
colissionless system was also recently rediscussed in the context
of this enlarged formalism \cite{lima}. In this latter study, it
was shown that such systems present instability even for
wavelenghts of the disturbance smaller than the standard critical
Jeans value. Recently, Taruya \& Sakagami \cite{Ta1,Ta2} discussed
a stability problem in the microcanonical ensemble by extending
Padmanabhan's \cite{pad} classical analysis of the Antonov
instability to the case of polytropic distributions functions.
%%%%%%%%%%
%In this work, since
%assuming the Tsallis' entropic function as the generalization from
%the Gibbs-Boltzmann's, the authors have worked with the
%extremum-entropy states under the constrains from the energy and
%normalization.
%%%%%%%%%
More recently, Chavanis \cite{chav1,chav2,chav3} considered the
Tsallis' entropy as a particular $H$-function corresponding to
isothermal stellar systems and stellar polytropes. In this
approach, the maximization of the $H$-function at fixed mass and
energy reveals naturally a thermodynamical analogy with the study
of the dynamical stability in collionless stellar systems
\footnote{For a complete and updated list of references see
http.tsallis.cat.cbpf.br/biblio.htm}.

In this {\it Letter}, we discuss the negative nature of the
specific capacity $C_V$ of self-gravitating systems in the
framework of a kinetic theory resulting from this non-extensive
formalism. To do so, we consider self-gravitating systems that can be treated as 
isolated systems in a first approximation, e.g., globular cluster \cite{LB} or elliptical 
galaxies \cite{HM}. We derive a new analytical expression for this quantity
which gives rise to an entire negative branch. We show that for this kind of system 
the value $q = 5/3$ is an upper limit for the non-extensive parameter. 
In connection with stellar
polytropes, the value of the non-extensive parameter
$q=5/3$ corresponding to a polytropic index $n=3$ or $n= -1$, depending on the expression between 
$n$ and $q$.

It has been shown that the kinetic foundations of the Tsallis'
themostatistic lead to a velocity distribution for
free particles given by \cite{rai,lima1}
\begin{equation}\label{fq}
f_0(v)=B_q \left[1-(1-q){mv^2\over 2k_B T}\right]^{1/1-q}.
\end{equation}
In this velocity $q$-distribution, the $q$-parameter
quantifies the nonadditivity property of the associated gas
entropy whose the main effect at the level of the distribuition is
to replace the standard Gaussian form by a power law.
Mathematically, this result follows directly from the well known
identity $\rm{lim}_{w\rightarrow 0}(1+wx)^{1/w}=\exp(x)$
\cite{Abram}. The quantity $B_q$ is a $q$-dependent
normalization constant whose expression for $ q \leq 1$ and $ q
\geq 1$ are respectively given by
\begin{eqnarray}\label{eq13}
{B_{q \leq 1} \over B_{q \geq 1}} &=&\left\{\begin{array}{ll}
n(1-q)^{1/2}(\frac{5-3q}{2})(\frac{3-q}{2})\frac{\Gamma
(\frac{1}{2}+{1\over
1-q})}{\Gamma({1\over 1-q})}\left(\frac{m}{2\pi k_BT}\right)^{3/2} \\
n(q-1)^{3/2}\frac{\Gamma({1\over
q-1})}{\Gamma ({1\over q-1}-\frac{3}{2})}\left(\frac{m}{2\pi
k_BT}\right)^{3/2}
\end{array}\right.,
\end{eqnarray}
where $T$ is the temperature and $n$ is the particle number
density. This non-extensive velocity distribution can be derived
at least from two different methods: (i) through a simple
non-extensive generalization of the Maxwell ansatz, $f(v)\neq
f(v_x)f(v_y)f(v_z)$, which follows from the introduction of
statistical correlations between the components of the velocities
\cite{rai} and (ii) by using a more rigorous
treatment based on the non-extensive formulation of the Boltzmann
$H$-theorem, which requires $q > 0$ \cite{lima1}.

In order to investigate the $q$-dependence of the specific
capacity, we first consider a cloud of ideal gas within the
nonrelativistic gravitational context, which is analog to a
self-gravitating collissionless gas. The kinetic energy of this
system is simply given by $K={1\over 2}m<v^2>$, with the
statistical content included in the average square velocity of the
particles $<v^{2}>$. In this way, the non-extensivity is introduced through
a new derivation of expectation value \cite{tsallis1}
\begin{equation}
<v^2>_q = {\int_{-v_m}^{v_m} f^q v^2 d^3 v\over\int_{-v_m}^{v_m} f^q
d^3 v},
\end{equation}
where $v_m=\left({2k_B T\over m(1-q)}\right)^{1/2}$ is a thermal
cutoff on the maximum value allowed for the velocity of the
particles ($q<1$), whereas for the power law $q$-distribution
without cutoff ($q>1$) $v_m \rightarrow \infty$. This
$q$-expectation value can be easily evaluated for $q\neq 1$, and
the result corresponds to a $q$-parameterized family of the square
velocity of the particles $<v^2>_q$ given by
\begin{equation}\label{9}
<v^2>_q={6\over 5-3q} {k_B T\over m}, \quad   \mbox{for $q < 5/3$},
\end{equation}
with $<v^2>={3 k_B T\over m}$ being the classical value easily
obtained in the limit $q \rightarrow 1$ (It is
worth mentioning that the virial theorem is not modified in this
non-extensive formalism \cite{mart}).

\begin{figure}
\vspace{.2in}
\centerline{\psfig{figure=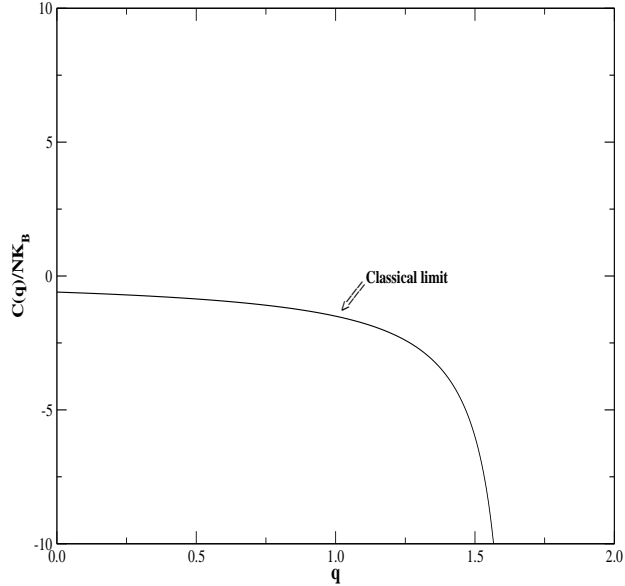,width=3.3truein,height=3.3truein,angle=-90}
\hskip 0.1in} \caption{The quantity $C_V/Nk_B$ for a
self-gravitating collisionless gas obeying the non-extensive
Tsallis' $q$-statistic is shown as a function of the non-extensive
parameter $q$. We see that there is a negative branch for the interval $0< q < 5/3$. 
For $q=5/3$ the
specific capacity diverges ($C_V\rightarrow - \infty$).}
\end{figure}

Now, by substituting the non-extensive $q$-expectation value given
by Eq. (\ref{9}) into the heat capacity relation Eq. (\ref{1}) one
finds
\begin{equation}
C_V=-{3\over 5-3q} Nk_B,
\end{equation}
which clearly reduces to the Maxwellian limit $C_V=-{3\over
2}Nk_B$ for $q = 1$. To better understand the physical
implications of the above expression, in Fig. 1 we plot the
quantity $C_V/Nk_B$ as a function of the non-extensive parameter
$q$. The graph clearly shows that the specific capacity is a negative quantity for the interval 
$q < 5/3$. For $q =
5/3$ (critical value) the specific capacity diverges
($C_V\rightarrow - \infty$). From this analysis, it is possible to
conclude that this particular $q$-dependence of $C_V$ leads to the
classical gravothermal catastrophe ($C_V<0$) only for power laws
with non-extensive parameter in the range $0 < q < 5/3$.

At this point we emphasize that the true nature of the
non-extensive $q$-parameter appearing in the Tsallis' statistical
framework remains as a completely open question at present. In
this way, it is important to show empirical manifestations of
these non-extensive effects. In this {\emph {Letter}} we have
discussed a clear signature of non-extensive effects on some kind of 
self-gravitating systems. In particular, we have shown that the $q$-dependence of
the specific capacity for these systems is physically consistent only for 
power-laws lying in the interval $0<q<5/3$. This particular range of values for the non-extensive
parameter $q$ is less restrictive than that one obtained by Abe
\cite{abe}. We suspect that such a difference occurs because the Abe's approach is addressed in the context of the canonical ensemble and, as has been shown elsewhere, in this ensemble self-gravitating systems do not
present negative specific heat \cite{pad}. We also note that, when
identified with the polytropic index $n = {3\over 2}+{1\over q - 1}$
\cite{plast}, the critical 
value $q=5/3$ is identical to the one
that gives rises to the gravitational stability/instability
transition, i.e., $n = 3$ \cite{chav1} (or $n = -1$ if $n = {1\over 2}+{1\over 1 - q}$ 
\cite{Ta2}). In a
couple of papers by Taruya \& Sakagami \cite{Ta2,Ta3} it was shown
that a thermodynamic instability appears for values of $n>5$ in
systems confined in adiabatic walls (microcanonical ensemble) and
for values of $n>3$ in systems surrounded by a thermal bath. In fact, a more consistent study of the gravothermal catastrophe must consider a self-gravitating ideal gas contained in a spherical container. We shall return to this discussion in a more complete analysis involving general non-extensive kinetic theories which will appear
in a forthcoming communication.

\vspace{0.5cm}

{\it Acknowledgments:} The authors are very grateful to J. A. S. Lima, P. H. Chavanis and A. 
Taruya for helpful discussions and a critical
reading of the manuscript. This work is supported by the project CNPq
(62.0053/01-1-PADCT III/Milenio). JSA is also supported by the
Conselho Nacional de Desenvolvimento Cient\'{\i}fico e
Tecnol\'ogico (CNPq - Brasil).

\end{document}